\documentclass[pra,aps,a4paper,showpacs,twocolumn]{revtex4}
\usepackage{amsmath,bm}
\usepackage{amsfonts}
\usepackage{amssymb}
\usepackage{graphicx}
\usepackage{dcolumn}
\usepackage{bm}

\begin{document}
\title{Directed coherent transport due to Bloch oscillation in two dimensions}


\author{J.~M. Zhang}
\affiliation{Beijing National Laboratory for Condensed Matter
Physics, Institute of Physics, Chinese Academy of Sciences, Beijing
100080, China} 
\affiliation{FOCUS Center and MCTP, Department of Physics, University of Michigan, Ann Arbor, Michigan 48109, USA} 
\author{W.~M. Liu}
\affiliation{Beijing National Laboratory for Condensed Matter
Physics, Institute of Physics, Chinese Academy of Sciences, Beijing
100080, China}

\begin{abstract}
We point out that in higher dimensions, in contrast to the one dimensional case considered usually, Bloch oscillation driven by a static force can induce transport of the wave packet. The wave packet oscillates constantly, but on a larger time scale it drifts at a constant velocity permanently. As a noteworthy feature, the net transport in the long run is always normal to the external force and thus controlled by it. We verify this prediction numerically and discuss its experimental realization both with cold atoms in optical lattices and with two dimensional photonic lattices. 
\end{abstract}
\pacs{03.75.Lm, 03.65.Ge, 03.65.Sq, 42.81.Qb} \maketitle

Bloch oscillation is a peculiar response of a particle in a periodic potential to an external force \cite{bloch}. Under a static uniform force, the particle performs an oscillatory motion in the real space without falling down to the potential minimum at infinity as in free space. This counter-intuitive phenomenon is purely quantum and has its root deep in the band structure associated with the periodic potential. The periodic potential mixes the plane waves into Bloch waves, which are classified by band number $n$ and wave vector $\bm{k}$. For a weak external force, the gap between the bands protects the particle from transition into other bands, with the only effect being that the wave vector is dragged across the Brillouin zone (BZ). The oscillation of the center-of-mass of the wave packet is then related to the fact that Bloch waves are periodic with respect to $\bm{k}$.

So far, Bloch oscillation has been investigated and observed in a variety of systems, such as semiconductor superlattice \cite{Waschke}, cold atoms in optical lattices \cite{dahan,kolovsky}, and photonic lattices \cite{pertsch,trompeter,longhi}. However, most of these works are confined to one dimension (some exceptions being \cite{kolovsky,trompeter}). In one dimension, the Brillouin zone $[-G/2,+G/2]$ has the topology of a circle since $-G/2$ is identified with $+G/2$. Therefore, in the ${k}$-space, the motion of the particle is simple---it transverses the Brillouin zone repeatedly. Quantitatively, we have the Newton second law type equation of motion \cite{kittel}:
\begin{equation}  \label{k eq 1}
 \frac{d {k}}{d t}={F}(t),
\end{equation}
where ${F}(t)$ is the time-dependent external force, and the wave vector ${k}$ is identified with ${k}+n {G}$ for an arbitrary integer $n$ ($\hbar=1$ in this paper). In the static case ${F}(t)\equiv {F}$, after a period of $T= G/F$, the wave vector returns to its initial value $k_0$ and so does the center-of-mass \cite{kittel,expl}:
\begin{eqnarray}\label{disp 1}
F \int_0^T dt \frac{d r}{d t} &=&F\int_0^T dt \nabla_k E(k_0+F t) \nonumber \\
&=& E(k_0+G)-E(k_0) =0. 
\end{eqnarray}
Here $E(k)$ is the dispersion function of the relevant energy band.
Therefore, in one dimension, Bloch oscillation induced by a static force does not lead to transport. To induce a net transport, a time-dependent force is needed. This makes sure that the Brillouin zone is transversed non-uniformly and thus the center-of-mass displacement accumulated in different $k$-regions do not cancel each other. This is essentially the scheme employed to induce the so-called super Bloch oscillations and macroscopic transport with cold atoms in one dimensional optical lattices \cite{thommen,alberti,haller}.

In this paper, we would like to point out that the situation is different in higher dimensions. In a higher dimensional potential, suppose the wave vector returns ever to its initial value by been dragged by some reciprocal lattice vector $\bm{G} $, the counterpart (or generalization) of Eq.~(\ref{disp 1}) is
\begin{eqnarray}\label{disp 2}
\bm{F} \cdot \int_0^T dt  \frac{d \bm{r}}{d t} &=&  \int_0^T d(\bm{F}t) \cdot  \nabla_{\bm{k}} E(\bm{k}_0+\bm{F} t) \nonumber \\
&=& E(\bm{k}_0+\bm{G})-E(\bm{k}_0) =0. 
\end{eqnarray}
Thus we can only arrive at the conclusion that as the wave vector returns to its initial value, the displacement of the center-of-mass of the wave packet must be perpendicular to the external force. But it does not necessarily vanish.

In the following, we give an explicit expression of the displacement in one period. We take the tight binding model to consider the Bloch oscillation. Note that, in a real lattice, Bloch oscillation is a good picture only when the single-band approximation is valid, i.e., when the band-band transition or the Zener tunnelling is negligible. With a tight binding model on a simple Bravais lattice, there is one and only one band, and the system is then free of band-band transition. Conversely, if we restrict to a single  band, then effectively we are dealing with a tight binding model on a simple lattice. The Wannier functions associated with that band in each primitive cell serve as the basis states on each site. Of course, by restricting to a single band, we necessarily misses the potential Berry phase effect and the transport associated \cite{niu}. However, since generally the Berry curvature is small in most area of the BZ \cite{yao}, or even vanishes identically in the presence of both time reversal and spatial inversion symmetry \cite{niu,as}, we neglect it in this paper.

One advantage of shaping the problem on a lattice is that, quantities of only a metric value are irrelevant. To be precise, the exact shape of the lattice, e.g. the lengths of the two basis vectors and the angle between them are of no concern. We then simply take the two basis vectors along the $x$ and $y$ directions, and assume their lengths to be unity. The first Brillouin zone is then $[-\pi,\pi]\times [-\pi,\pi]$. The original real lattice differs from this square lattice by only an affine transform.

The tight binding Hamiltonian is
$ H_{TB}=-\sum_{\bm{l},\bm{m}} J_{\bm{m}} |\bm{l}\rangle \langle \bm{l}+\bm{m} |$.
Here $\bm{l}$ and $\bm{m}$ both take values among all the integer
pairs $(m_1,m_2)$, $-\infty < m_{1,2} < \infty$, and $|\bm{l}\rangle$ denotes the Wannier funtion at site $\bm{l}$. Because
$H_{TB}=H_{TB}^\dagger$, we have $J_{\bm{m}}=J_{-\bm{m}}^*$. The Bloch states are $\Psi_{\bm{k}}=\sum_{\bm{m}} e^{i \bm{k} \cdot
    \bm{m}} |\bm{m}\rangle $, where $\bm{k}=(k_1,k_2)$ is the wave vector. They are eigenstates of $H_{TB}$ with eigenvalues
$ E(\bm{k})=- \sum\nolimits_{\bm{m}} J_{\bm{m}} \exp(i \bm{k} \cdot
    \bm{m})$.
In many cases, the original Hamiltonian is time reversal invariant, and therefore $E(\bm{k})=E(-\bm{k})$ necessarily. This implies $J_{\bm{m}}=J_{-\bm{m}}$. Combined with the previous condition, this implies $J_{\bm{m}}$ is real.
    
Now suppose a linear potential is applied to the system. The Hamiltonian is now
$ H=H_{TB}-\sum_{\bm{m}} \bm{F}\cdot \bm{m}  |\bm{m}\rangle \langle \bm{m} |$.
In the semi-classical theory, the details of the wave packet are neglected and it is characterized
just by a pair of variables $(\bm{r},\bm{k})$, which are respectively the average values of the position and moment of the wave packet. The semi-classical
equations of motion are \cite{kittel}
\begin{eqnarray}
  \frac{d \bm{k}}{d t} &=& \bm{F}, \label{k motion} \\
  \frac{d \bm{r}}{d t} &=& \nabla E(\bm{k})=-i \sum\nolimits_{\bm{m}}  \bm{m}
  J_{\bm{m}}
\exp(i \bm{k} \cdot
    \bm{m}). \label{r motion}
\end{eqnarray}
These equations can be understood as follows. First note that an initial Bloch state remains as a Bloch state all the time, more precisely, the time varying Bloch state $\Psi_{\bm{k}_0+\bm{F}t} e^{-i\int_0^t d t_1 E(\bm{k}_0+\bm{F} t_1)}$ solves the Schr\"odinger equation $i \partial\psi /\partial t=H \psi$. This actually implies or solves Eq.~(\ref{k motion}). Now for an initial wave function
$\psi_{\bm{m}}(0)=\int_{BZ} d \bm{k} f (\bm{k}) e^{i \bm{k} \cdot \bm{m}}$,
where $f(\bm{k})=f(\bm{k}_0 +\delta \bm{k})\equiv g(\delta \bm{k})$ is a function narrowly localized around $\bm{k}_0$ (so that it is meaningful to say the wave vector of the initial wave function is $\bm{k}_0$), the wave packet at time $t$ is 
\begin{eqnarray} \label{psi t}
\psi_{\bm{m}}(t) &=&\int_{BZ} d \bm{k} f (\bm{k}) e^{i (\bm{k}+\bm{F} t) \cdot \bm{m}}e^{-i\int_0^t d t_1 E(\bm{k}+\bm{F} t_1)} \nonumber \\
&\simeq & \int d (\delta\bm{k}) g (\delta \bm{k}) e^{i \delta \bm{k} \cdot (\bm{m}-\int_0^t d t_1 \nabla E(\bm{k}_0+\bm{F} t_1))} \nonumber \\
&& \quad  \times e^{i (\bm{k}_0+\bm{F} t) \cdot \bm{m}}e^{-i\int_0^t d t_1 E(\bm{k}_0+\bm{F} t_1)}.
\end{eqnarray}
Here the phase factor in the first line is expanded around $\bm{k}_0$. This is legitimate because the main contribution to the integral comes from a small neighborhood of $\bm{k}_0$. The second line implies that the wave packet is translated forward by
\begin{eqnarray} \label{dd}
  \bm{r}(t)-\bm{r}(0) &=& \int_0^t d t_1 \nabla E(\bm{k}_0+\bm{F} t_1) \nonumber \\
  &=& -i \sum\nolimits_{\bm{m}}  \bm{m}
  J_{\bm{m}} \int_0^t d t_1 
e^{i (\bm{k}_0 \cdot
    \bm{m}+ \bm{F}\cdot \bm{m} t_1)},\quad
\end{eqnarray}
which solves Eq. (\ref{r motion}).

Assume that there exist some time $T$ such that $\bm{F} T$ is equal to a reciprocal lattice vector, that is,
\begin{equation}\label{ft}
\bm{F} T=(F_1 T, F_2 T)=( 2\pi q , 2\pi r),
\end{equation}
where $q$ and $r$ are some co-prime integers. For $F_1/F_2=q/r$, the explicit value
of $T$ is $T=2\pi q/F_1$. After one period of $T$, the wave vector
returns to its initial value since $\bm{k}$ is identified with
$\bm{k}+( 2\pi q , 2\pi r)$. The displacement
of the wave packet in this period is calculated as
\begin{eqnarray} \label{disp 10}
\bm{D}_{T} &=& -i \sum\nolimits_{\bm{m}}  \bm{m}
  J_{\bm{m}} \int_0^T d t_1 
e^{i (\bm{k}_0 \cdot
    \bm{m}+ \bm{F}\cdot \bm{m} t_1)} \nonumber \\
    &=& T \sum\nolimits_{\bm{m}} \bm{m}
  J_{\bm{m}} \sin(\bm{m} \cdot\bm{k}_0) \delta_{\bm{F} \cdot \bm{m}}.
\end{eqnarray}
This equation is our central result. The Dirac function means that for a given $\bm{F}$ in some direction, only those $\bm{m}$ 
which are perpendicular to $\bm{F}$ contribute to the net displacement.
Physically, this means in the long run the wave packet drifts
perpendicular to the external force. This is quite reasonable. If
after one period of $T$, the wave packet is shifted somewhat along
the direction of the external force, then in the long run, the shift
of the wave packet along the direction of the external force will
diverge to $+\infty$ or $-\infty$ linearly with time. As the system
is conservative, the kinetic energy of the particle will go to $+\infty$
or $-\infty$ respectively. However, this is
impossible since in a definite band, $E(\bm{k})$ has both upper
and lower bounds. The argument applies equally well to any
dimensions and holds as long as the single band approximation is
valid. 

Another remark is worthy. Our assumption (\ref{ft}) about the direction of $\bm{F}$ seems quite stringent. Indeed, since the measure of rational numbers is zero, for a generic $\bm{F}$, the ratio $F_1/F_2$ 
is irrational and thus Eqs.~(\ref{ft}) and (\ref{disp 10}) seem irrelevant. Moreover, even restricted to the field of rational numbers, $T$ and $\bm{D}_{T}$ are pathologically sensitive to the ratio $F_1/F_2$ (e.g., suppose $F_1/F_2$ is perturbed from $1/2$ to $101/200$, $T$ enlarges by a factor of $100$ while $\bm{D}_{T}$ shrinks almost to zero since the associated hopping coefficients are exponentially small). The way out of this anomaly is that $T$ and $\bm{D}_T$ both refer to the overall behavior in the long run and thus they become irrelevant if $T\gg\tau$, the observation time span. We further note that in a given finite time interval, the displacement $\bm{r}(t)-\bm{r}(0)$ in Eq.~(\ref{dd}) is a well behaved smooth function of $F_1/F_2$. Thus, as long as the motion in a given time interval is concerned, the concerns above are irrelevant. Of course, to make the directed transport effect as significant as possible, it is preferable to choose $\bm{F}$ in the vicinity of those directions $(q,r)$ with $q$ and $r$ not large. 

In the hindsight, we see from Eq. (\ref{psi t}) why Bloch oscillations in one dimension and in higher dimensions behave differently. In one dimension, all $\bm{k}$ vectors undergo the same trajectory and thus the phases accumulated are the same for all the $\bm{k}$ vectors. This means after one period, the wave function itself returns to its initial value (up to a global phase), not only its center-of-mass. In contrast, in higher dimensions, different $\bm{k}$ vectors may take different (parallel) trajectories in the $\bm{k}$-space, though they all return to their initial values simultaneously. The phases accumulated in one period for $\bm{k}$ vectors on different trajectories differ in general and this dispersion leads to the shift of the center-of-mass of the wave packet. 

\begin{figure*}[thb]
\begin{minipage}[b]{0.42 \textwidth}
\includegraphics[width=\textwidth]{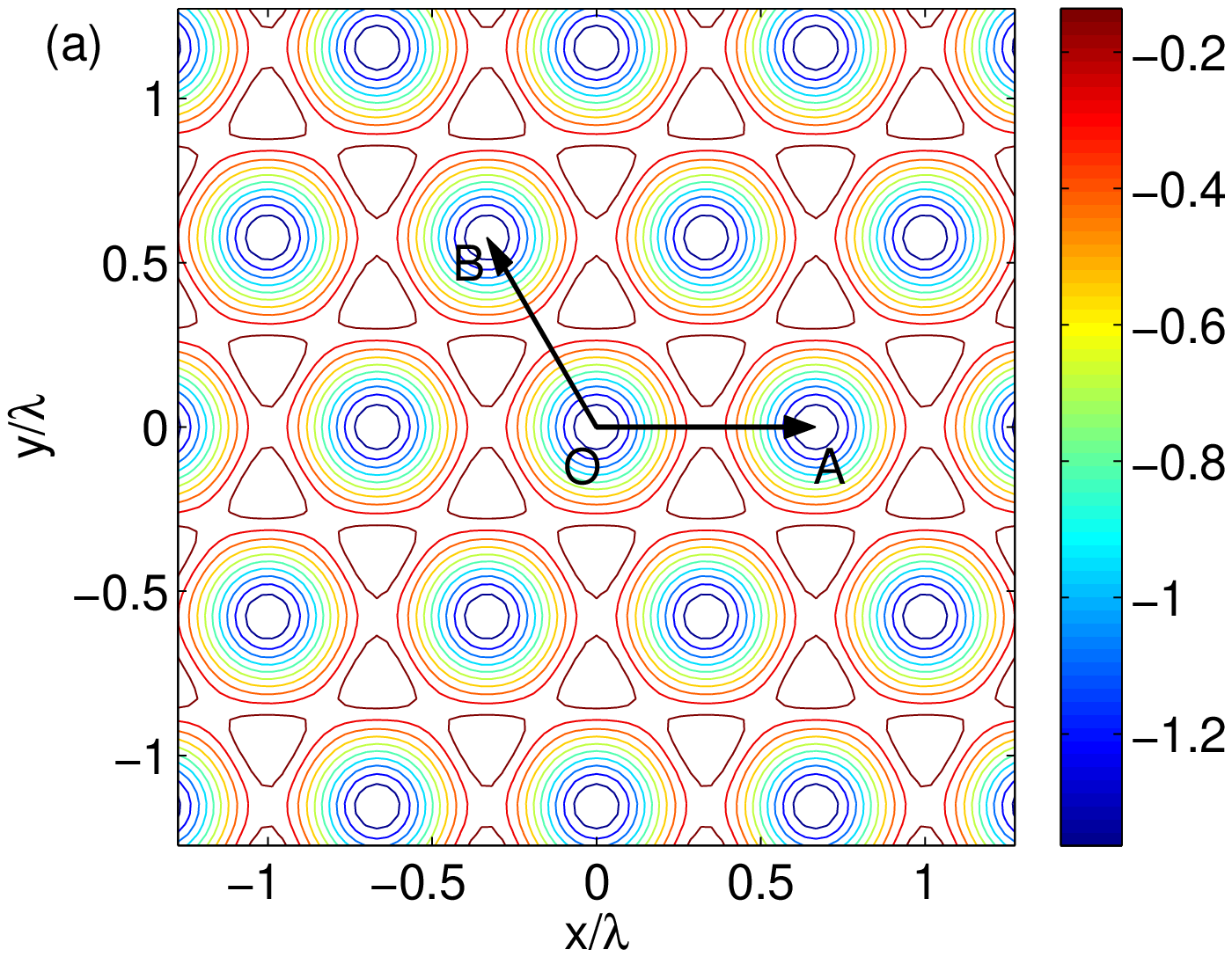}
\end{minipage}
\begin{minipage}[b]{0.42 \textwidth}
\includegraphics[width=\textwidth]{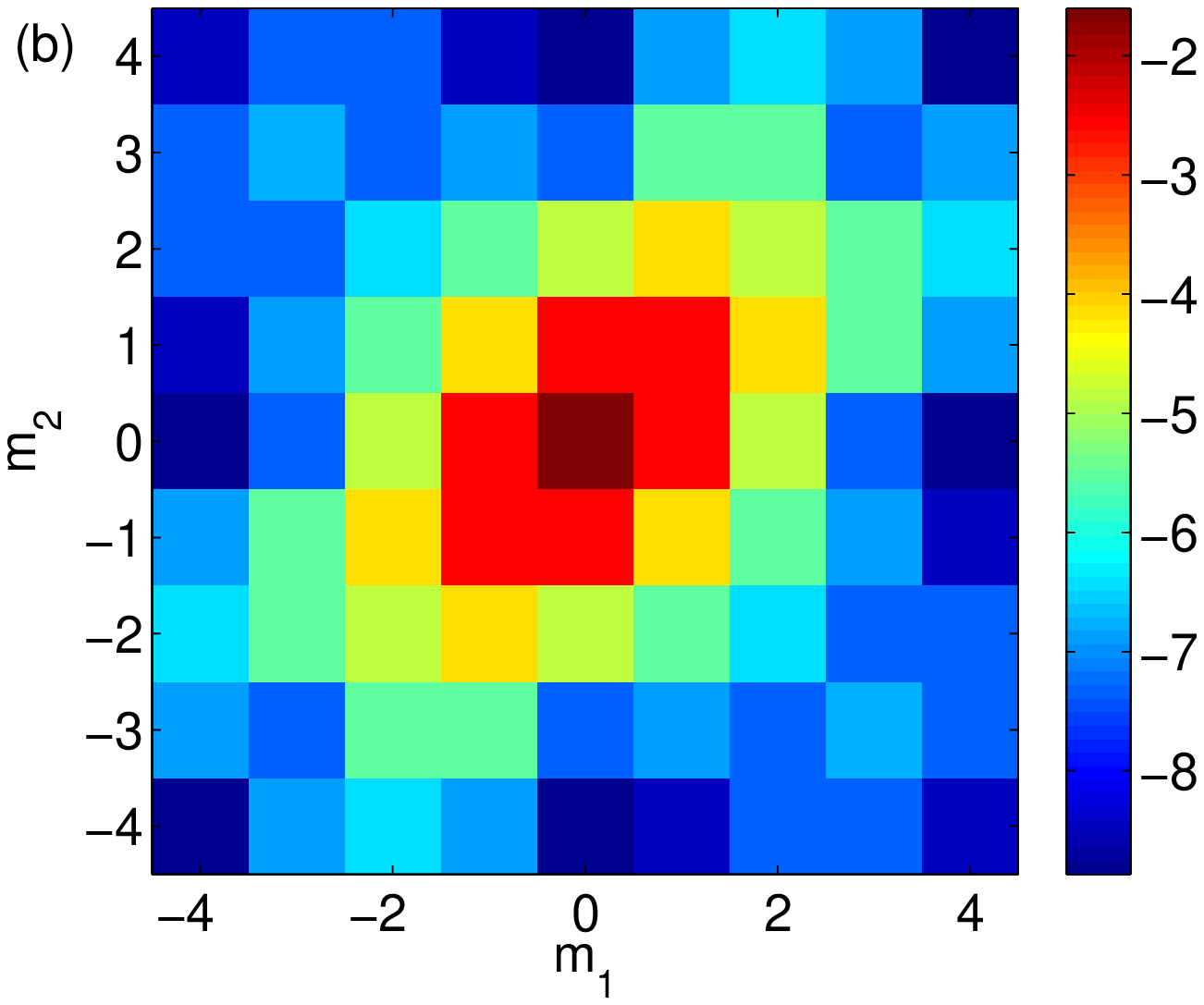}
\end{minipage}
\caption{\label{fig1}(Color online) (a) Contour plot of the triangular optical lattice with the minima $V_0=-1.5E_r$, with $E_r=h^2/2m\lambda^2$ being the recoil energy. The basis vectors chosen are $ OA $ and $OB$, with $|OA|=|OB|=2\lambda/3$. The coordinates of sites $A$ and $B$ are $(1,0)$ and $(0,1)$, respectively. (b) Natural logarithm of the absolute values of the hoppings $J_{(m_1,m_2)}$ in units of $E_r$. Note that the $D_6$ symmetry of the lattice in (a) is respected, e.g., we have exactly $J_{\pm(1,0)}=J_{\pm(0,1)}=J_{\pm(1,1)}\equiv J_1$.}
\end{figure*}

In the following we apply the formalism developed above to some concrete case. We consider the scenario of cold atoms in a planar triangular optical lattice. This lattice can be constructed by interfering three laser beams travelling $2\pi/3$ with respect to each other in the $xy$ plane and all polarized in the $z$ direction \cite{sengstock}. The laser beams are all of wave length $\lambda$ and amplitude $E$. Suppose the laser is red detuned from the atomic transition, then the minima (of value $V_0 \propto 9E^2$) of the optical potential form a triangular lattice (see Fig.~\ref{fig1}(a)). By solving the energy bands exact numerically and then performing Fourier transform by using the expression of $E(\bm{k})$, we can solve the hopping amplitudes $J_{\bm{m}}$ which are controlled by $V_0$. In Fig.~\ref{fig1}(b), we show the logarithm of their magnitudes $\ln|J_{\bm{m}}|$ in the case of $V_0=-1.5E_r$, with $E_r=h^2 /2m\lambda^2 $ being the recoil energy. We note that the site-site hopping decreases monotonically with the site-site distance. Moreover, the $D_6$ point group symmetry of the lattice is respected perfectly. In our numerical simulations below, we will only preserve site-site hoppings belonging to the first three hopping values $J_i$ $(i=1,2,3)$ according to their magnitudes. In the specific case of $V_0=-1.5E_r$, $(J_1,J_2,J_3)=(0.0765,-0.0149,-0.0078) E_r$. Those site-site hoppings belonging to $J_1$ are $J_{\pm (1,0)}$, $J_{\pm (0,1)}$, and $J_{\pm (1,1)}$, and those belonging to $J_2$ are $J_{\pm (2,1)}$, $J_{\pm (1,2)}$, and $J_{\pm (-1,1)}$, and to $J_3$ are $J_{\pm (2,0)}$, $J_{\pm (0,2)}$, and $J_{\pm (2,2)}$.

\begin{figure*}[thb]
\begin{minipage}[b]{0.45 \textwidth}
\includegraphics[width=\textwidth]{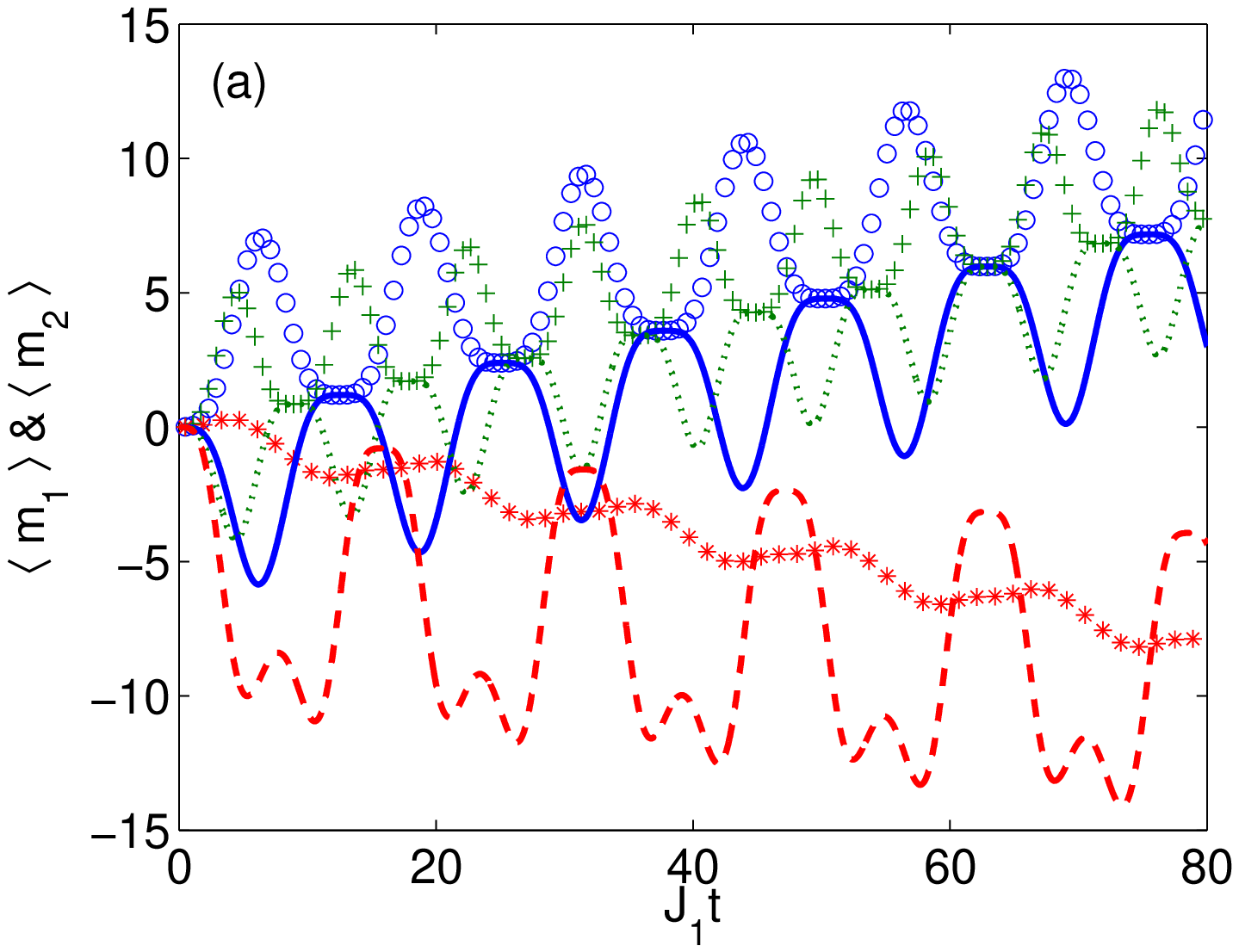}
\end{minipage}
\begin{minipage}[b]{0.45 \textwidth}
\includegraphics[width=\textwidth]{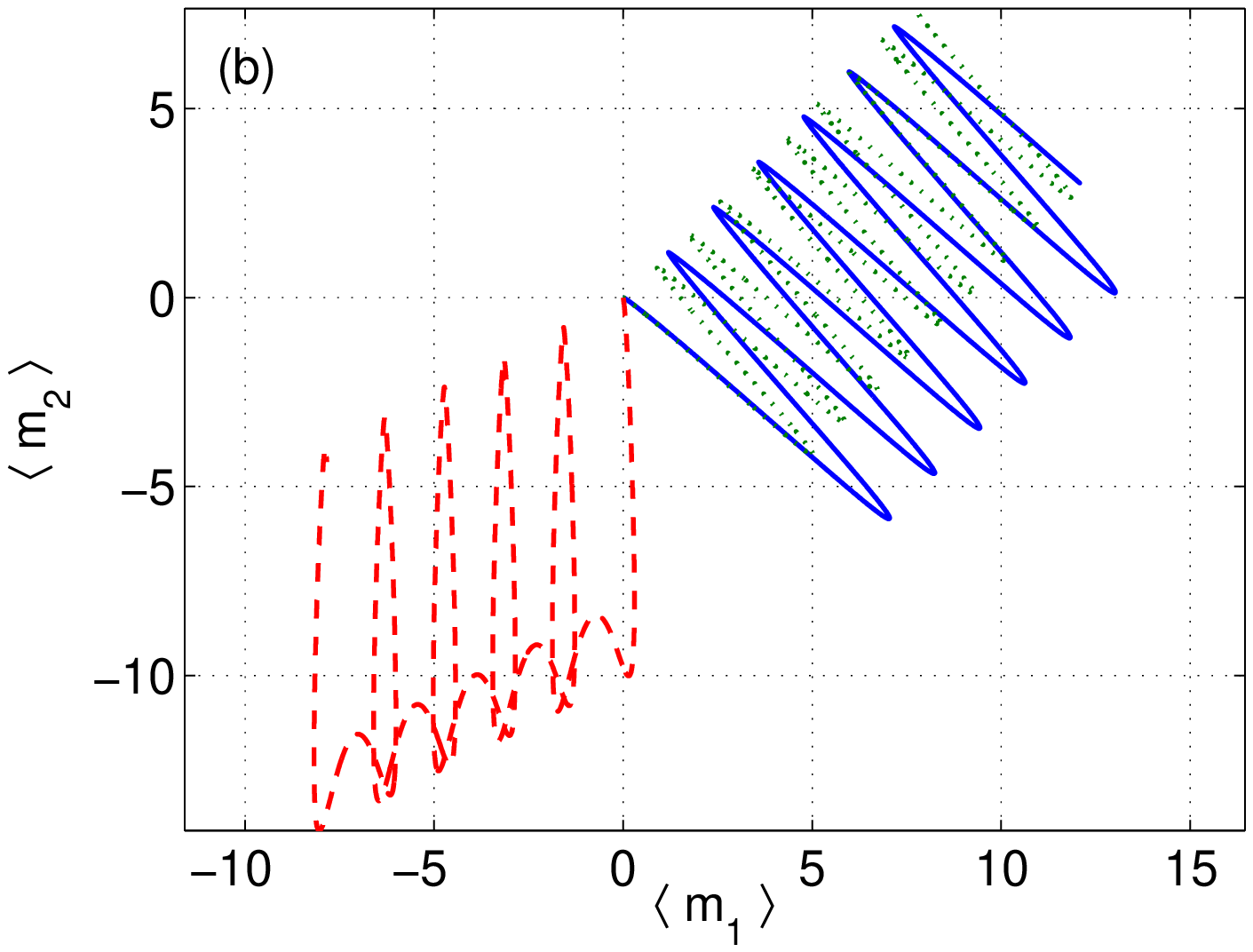}
\end{minipage}
\caption{\label{fig2}(Color online) (a) Time evolutions of the center-of-mass $\langle m_1 \rangle$ and $\langle m_2 \rangle$ of a wave packet in the optical lattice depicted in Fig.~\ref{fig1}. The initial wave packet is a Gaussian $\psi_{\bm{m}}(t=0)=A\exp(-(m_1^2+m_2^2)/\sigma^2+i k_0^1 m_1+i k_0^2 m_2)$, with width $\sigma=20$ and wave vector $\bm{k}_0=(k_0^1,k_0^2)=(0.05,0.03)$. Evolution of the wave packet under three different forces are considered. They are (i) $\bm{F}/J_1=(0.5,-0.5)$, (ii) $\bm{F}/J_1=(0.7,-0.7)$, and (iii) $\bm{F}/J_1=(0.4,-0.8)$, respectively. In each case we have respectively, (i) $\langle m_1 \rangle$: blue $\circ\circ\circ$ line, $\langle m_2 \rangle$: blue solid line; (ii) $\langle m_1 \rangle$: green ${+++}$ line, $\langle m_2 \rangle$: green dotted line; (iii) $\langle m_1 \rangle$: red $\ast\ast\ast$ line, $\langle m_2 \rangle$: red dashed line. (b) Trajectories $\langle m_1 \rangle$ versus $\langle m_2 \rangle$ in the three cases. The solid, dotted, and dashed lines refer to the cases (i), (ii), and (iii), respectively.}
\end{figure*}

The initial wave packet is assumed to be a Gaussian, $\psi_{\bm{m}}(t=0)=A\exp(-(m_1^2+m_2^2)/\sigma^2+i k_0^1 m_1+i k_0^2 m_2)$, with $\bm{m}=(m_1,m_2)$. Here $\sigma=20$ is the width of the wave packet and $\bm{k}_0=(k_0^1,k_0^2)=(0.05,0.03)$ is the initial wave vector, while $A$ is some normalization factor. The width $\sigma\gg 1$ so that the wave function is well localized around $\bm{k}_0$ in the $\bm{k}$-space. Three cases of different forces are investigated. They are (i) $\bm{F}/J_1=(0.5,-0.5)$, (ii) $\bm{F}/J_1=(0.7,-0.7)$, and (iii) $\bm{F}/J_1=(0.4,-0.8)$, respectively. The  Schr\"odinger equation $i\partial \psi/\partial t=H \psi$ is solved on a $121 \times 121$ lattice by the fourth order Runge-Kutta method. The time evolutions of the center-of-mass of the wave packet $\langle m_1 \rangle$ and $\langle m_2 \rangle$, are shown in Fig.~\ref{fig2}(a). As another respect, the trajectories $\langle m_1 \rangle$ versus $\langle m_2 \rangle$ are shown in Fig.~\ref{fig2}(b). The semi-classical predictions according to Eqs. (\ref{k motion}) and (\ref{r motion}) agree with the exact numerical calculations perfectly and thus are not shown. In all the cases, we see that though there are temporary oscillatory deviations, in the long run, the wave packet drifts at a constant velocity perpendicular to the external force. In cases (i) and (ii), the drifting direction is along $(1,1)$, while in case (iii), it is along $(2,1)$, as evident in Fig.~\ref{fig2}(b). We also note, by comparing case (i) with case (ii), that the oscillation amplitude is suppressed for a larger force, and on the other hand, the oscillation period is shortened. These features are reminiscent of Bloch oscillation in one dimension. For a larger force, the Brillouin zone is transversed faster and thus the period is shortened, and so is the oscillation amplitude because the deviation accumulated decreases accordingly. The three cases show how the transport direction and the detailed temporary motion of the wave packet can be controlled by the direction and magnitude of the external force, respectively.

In conclusion, we have demonstrated that Bloch oscillations in higher dimensions, in contrast to their counterpart in one dimension, can lead to transport of the wave packet. Moreover, the direction of the transport is always perpendicular to and is thus controlled by the external force. In view of the intensive investigations on the control of transport of ultracold atoms in optical lattices \cite{alberti,haller,sias}, this fact may find use in future experiments. As for the experimental observation of the scenario considered in this paper, we would say that the system can be readily realized with current experimental technology. The optical lattice can be constructed routinely \cite{sengstock} and the external force can be provided by the gravity \cite{anderson}. By tilting the two dimensional optical lattice from the horizontal plane, both the magnitude and the direction of the force can be adjusted at will. Furthermore, the great progress in \textit{in-situ} observation of cold atoms in real space \cite{wurtz} guarantees that observation of the directed transport should also be achievable in the near future. Finally, we note that $1/J_1$ in Fig.~\ref{fig2}(a) corresponds to a time on the order of $0.6\sim 0.7$ ms for $^{87}$Rb, while the life time of cold atoms in an optical lattice can be made well on the order of 1 s \cite{alberti,haller}. Thus this system is long-lived enough to allow for a significant drift which is in turn readily observable experimentally. 

An alternative approach for experimental realization is by the two dimensional photonic array. There, the propagation of light in the waveguides maps to the tight binding model perfectly, and the force can be effected by engineered refractive index variation \cite{longhi}. Actually, two dimensional Bloch oscillation and Zener tunnelling have been observed in this system \cite{trompeter}. Therefore, it is likely that directed transport should also be observable in this system, which would show up as drift of the center of the pattern with the propagation length.

This work was supported by NSF of China under Grant Nos.~10874235, 10934010, and 60978019, NKBRSF of China under Grant Nos. 2006CB921400, 2009CB930701, and 2010CB922904.


\begin{thebibliography}{99}
\bibitem{expl}
Note that when talking about Bloch oscillation, we have in mind a (localized but sufficiently wide) wave packet not an (infinitely extended) exact Bloch state. Therefore, the wave vector $k$ and the position $r$ all refer to their average values. In the first line, the standard expression of the packet's group velocity is used. See the paragraph containing Eq.~(\ref{psi t}) for more explanations.
\bibitem{as}
As is the case with the triangular optical lattice in this paper.
\end{thebibliography}
\end{document}